# Reliability constrained least-cost generation expansion planning: an isolated mini-grid in KSA

**Markus Groissböck [1,a,b], Alexandre Gusmao [2,c]**

[1] markus.groissboeck@student.uibk.ac.at, https://orcid.org/0000-0001-6025-9477
[2] alexander.gusmao@aramco.com
[a] Saudi Aramco, Power Systems Department, Dhahran, Kingdom of Saudi Arabia
[b] University of Innsbruck, Institute for Construction and Materials Science, Innsbruck, Austria
[c] Saudi Aramco, Power Systems Renewable Department, Dhahran, Kingdom of Saudi Arabia

***Abstract* — Renewable energy sources (RES) are getting closer to grid parity within interconnected systems while their use in isolated and remote areas is already cost competitive with conventional generation. In mini-grids, isolated and far from the grid, the economics of different technology options (e.g. pure solar photovoltaic (PV), wind, concentrated solar power (CSP), and open cycle gas turbine (OCGT)) with different amounts of added capacity is assessed to allow the peak demand to increase by 5 MW (or 5%) and 10 MW (or 10%). This assessment is based on keeping the system reliability (measured in loss of load expectation, LOLE) at the same level as it was before the addition of conventional and renewable generators. An effective load carrying capability (ELCC) increase by 5 MW (or 5%) can be realized by adding a 5 MW OCGT or 10 MW PV and 10 MW wind. An ELCC increase of 10 MW (or 10%) requires either 11 MW OCGT or 7.2 MW OCGT, 2.7 MW PV, and 8.1 MW wind. In both cases, the solution with RES has 8% and 4% lower annual costs than the OCGT only solution (assuming crude costs of $40/barrel). From an economic and system reliability perspective, systems with low growth rates (<5%) in peak and energy demand could focus on adding RES capacity while systems with higher growth rates (>5%) still need to incorporate conventional generation.**

## NOMENCLATURE

CC .......... Capacity Credit
COPT ..... Capacity Outage Probability Table
CSP......... Concentrated Solar Power
DR .......... Demand Response
DSM ....... Demand Side Management
ED .......... Economic Dispatch
ELCC ..... Effective Load Carrying Capability
EENS...... Expected Energy Not Served
FOR ........ Forced Outage Rate
GEP ........ Generation Expansion Plan
KSA ....... Kingdom of Saudi Arabia
LDC ....... Load Duration Curve
LOLE ..... Loss of Load Expectation
LOLP ..... Loss of Load Probability
OCGT .... Open Cycle Gas Turbine (~SCGT)
PV .......... Photovoltaic
RES ........ Renewable Energy Sources
SCGT ..... Simple Cycle Gas Turbine (~OCGT)
VER ....... Variable Energy Resources

## I. INTRODUCTION

Renewable energy sources (RES) are getting more and more competitive for interconnected grid usage. RES tends to be even more competitive in isolated and far from the grid locations as building transmission lines and fuel transportation can be very expensive. Generation expansion planning (GEP) in its nature is a nonlinear and dynamic problem to be solved under given restrictions and constraints [1]. In general, the purpose of GEP is to minimize required investments for generation and operational costs, including fuel costs to maximize the power plant owner's profit under a defined reliability metric target. This is true, especially for isolated areas (as used in this case study), as building transmission lines to connect them to the bulk system requires significant expense. The role of RES in mitigating fuel consumption in isolated areas is well-known as isolated areas usually use high-value hydrocarbon-based fuels with an extra premium coming from fuel transportation, which is in many cases difficult and expensive. Nevertheless, the other role RES has is contributing to the peak demand fulfillment of the power system. The peak demand fulfillment is measured by the effective load carrying capability (ELCC) of the RES. It is known that the ELCC from RES lessens with its higher penetration of the system's capacity. Therefore, it is important to understand the cost compromise between the savings on fuels with increasing penetration and the ELCC of the additional capacity added to the power system whenever the system serves a growing power load, but always under the same reliability target [2].

Consequently, this work contributes to the state-of-the-art ELCC considerations in a way that the probabilistic approach of loss of load expectation (LOLE) is considered in a dynamic programming framework in which a reliability constrained least-cost GEP is done to achieve a required peak increase of 5 and 10 %, respectively. The studied isolated mini-grid lies within the southwest of the Kingdom of Saudi Arabia (KSA) and will be further on referenced as the mini-grid. The reliability of each hour is based on loss of load probability (LOLP) to calculate the LOLE for an entire year. Within this work, a set of technology options will be ranked in terms of economics allowing the existing mini-grid to raise the peak demand while keeping the reliability at least constant and minimize the expected total life cycle costs, i.e. investment costs as well as operational expenditures including fuel costs.

The structure of the remaining work is as follows: Chapter II contains a literature review about different reliability criteria and best practices in systems studies. Chapter III contains the methodology discussing the least-cost generation expansion planning considering a dynamic programming approach. Chapter IV presents the considered case study based on an isolated mini-grid in KSA. Chapter V contains a discussion and conclusion. Chapter VII contains possible future enhancements.

## II. Literature Review

The power system is characterized by a probabilistic generation model of the existing system via a capacity outage probability table (COPT) along with the base demand within the mini-grid, which will serve as a reference for the base system. The base demand is characterized by its measured hourly load time series, while the renewable time series are calculated based on satellite measurements and simulated with typical configurations. The convolution between COPT and demand defines the reference LOLE for an existing system [2]. Each generation fleet is unique as there is no system in the world in which the individual generator reliability is equally based on differences in how maintenance is done in addition to the differences in size and reliability. Therefore, each probabilistic generation model, characterized by its COPT, is unique. The initial task is to calculate a system's LOLE for a given generation fleet and the current peak load within the assessed power system [3]. After that, additional LOLE's are calculated assuming different peak loads (see the left curve in Figure 1). After adding a new generation, the system can fulfill a different and higher peak load as the original system was able to while fulfilling the same system reliability. The difference is the ELCC of the added generation technology. The ELCC is calculated by the difference in the horizontal axis of the two corresponding peak loads of the LOLE functions at the same reliability target (horizontal line at a determined LOLE level, labeled 'design criterion'). If this ELCC were taken as a percentage of the added new generation capacity, the result would be the capacity credit (CC).

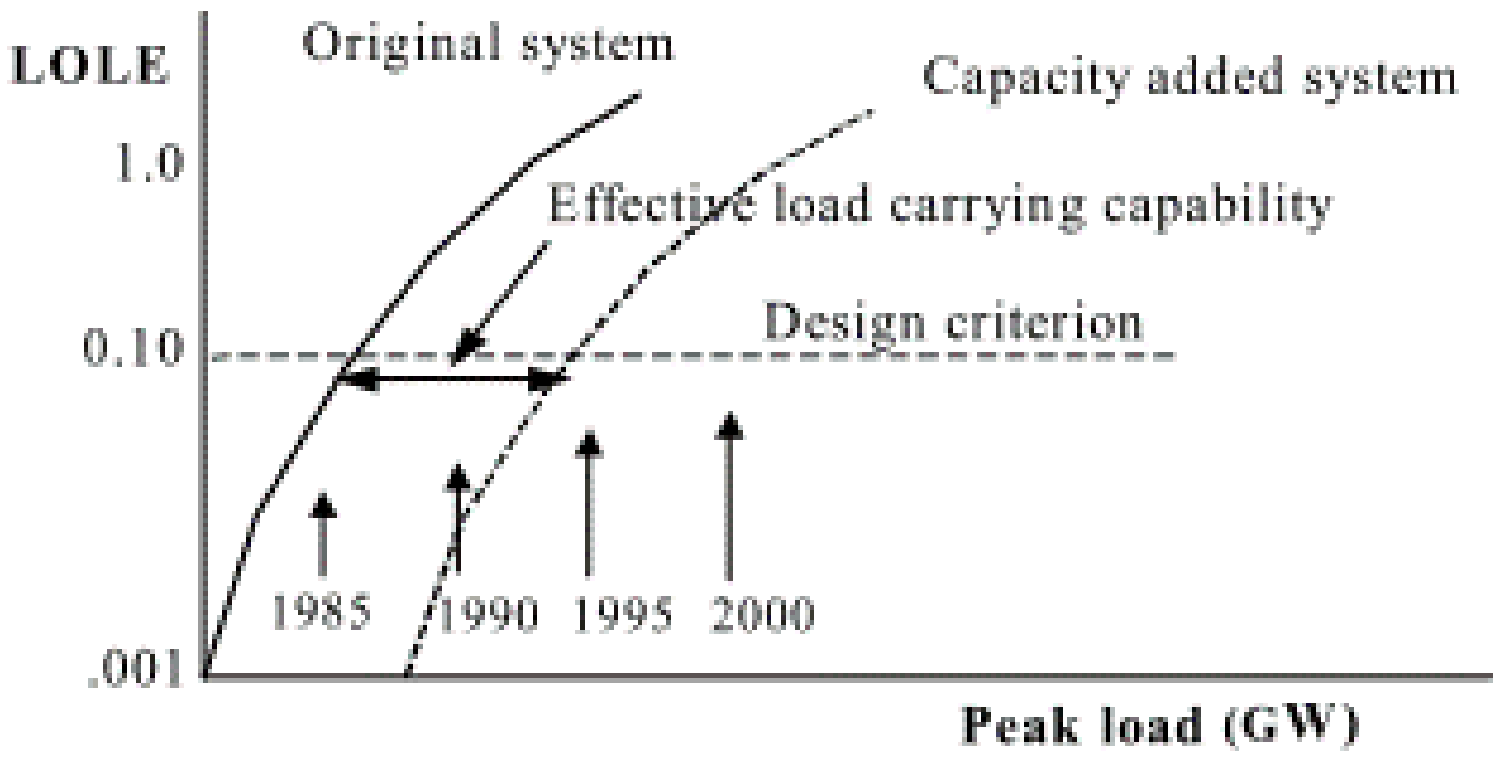

Figure 1: Effect of capacity additions on LOLE [3]

In the case of the presence of RES on the generation capacity expansion, net load should be considered as the input for the LOLE convolution instead of load [4]. Maintaining the current LOLE level, e.g. the power generation interruption risk level, is the core assumption behind probabilistic GEP. Figure 1 demonstrates how that can be shown in a graphical way. The ELCC represents the maximum increase in peak demand within the planning period at the same risk level as the previous system. It is important to mention that the term "existing system" is always a sum of the characteristics of the existing generators and incorporates the forced outage rate (FOR), and therefore the considered maintenance quality, the number and the size of the individual generators, as well as the shape of the load demand time series through the studied period [2].

In most of the cases, the expansion of an electric power system aims for finding the best alternative on the basis of total life time system costs [5]. Deterministic reliability metrics as reserve margin are not appropriate in an environment where more and more variability (demand response (DR), demand-side management, (DSM) and RES) is added into existing electricity systems. Variability is added through smart grid technologies as well as through an increasing amount of variable energy resources (VER) as solar photovoltaic (PV) and wind. Historically, GEP was mainly based on the deterministic performance indicator 'reserve margin'. The reserve margin defines how much the total installed power plant fleet should be above the expected peak demand to guaranty a reliable and secure power operation under all circumstances.

The total life time cycle costs of a power system should consider socioeconomic costs, emerging from the potential costs raised by society over the periods of non-served energy by the system. There are two main approaches to estimate the socioeconomic costs caused by interrupted power service: implicit and explicit [6]. In the explicit approach two points of views for the cost of interruptions are possible: i) point of view of the user, which implies that cost of interruptions may vary widely among consumers' type, and ii) point of view of the utility that may mean loss of revenue from load not served, potential loss of futures sales due to customer's reaction to low service reliability [5], or additional penalties from regulation authorities for reliability targets not being met. Common explicit criteria are e.g. expected energy not served (EENS, MWh/year) and loss of load frequency (LOLF, occurrences/year) [7]. While EENS represents the sum of the total energy the utility is not able to serve within a year, LOLF represents the number of hours a year the required energy cannot be supplied as requested. EENS does not provide any information on how often energy cannot be supplied. On the other hand, LOLF does not provide any information on how much energy cannot be provided within a specific period of time. The general disadvantage of this kind of explicit criteria is that both do not provide any insights on the probabilities of occurrence. Well-known tools like HOMER, ENERGYPLAN, and DER-CAM are not considering one of these explicit approaches [8] [9]. In the implicit approach, perhaps the most traditional, a comparison of various expansion alternatives is studied under the assumption that each of the alternatives is able to provide the same level of reliability criterion, meaning there is an implicit socioeconomic cost associated to the selected criterion. The implicit approach is used in this work. Variations on the reliability criterion mean different implicit socioeconomic cost of power interruptions. Therefore, to assume, for instance, a higher (lower) LOLE criterion for the expanded system implies a lower (higher) socioeconomic cost implication due to power interruptions. Within an isolated mini-grid it should be easier to establish an explicit socioeconomic cost, as the complexity is lower as for a whole nation. The socioeconomic cost should be part of GEP and therefore within the cost minimization process. This would incorporate the socioeconomic costs into all costs and therefore be part of the total system costs. This therefore would consider the expected non-served energy costs with the potential losses from social benefits as well as required investments, fuel and maintenance costs. Social benefits are, in its simplest way and ignoring different customer segments, measured as a fraction of the gross domestic product divided by the total energy demand throughout the year [10]. This is an easy approach to estimate the so-called value for the value of lost load (VOLL) indicator. A more accurate way of estimating VOLL would be based on customer's willingness to pay for reliable energy supply. And of course, different customer segments would have different VOLL estimates.

Historically, DER-CAM had not considered any explicit or implicit indicators [9]. Nevertheless, the latest publication shows an approach to consider N-1 contingencies based on penalties of not served electricity, which represents an implicit approach. Other tools such as the MATPOWER Optimal Scheduling Tool (MOST) can consider a multi-period security-constrained unit-commitment while incorporating e.g. ramp rates and volatile renewable generation [11] [12]. Also, this example shows that system reliability is not considered in a detailed and probabilistic way as LOLE should be. The computation burden is too high to be considered within a standard optimization framework.

In several studies in the U.S., sub-hourly time series data from multiple years has been considered to estimate the value of RES in terms of how much additional required conventional capacity can be reduced because of high RES additions [13] [14] [15]. Renewable integration studies in general consider multiple years of data for their RES and therefore ELCC is not necessary to be considered. Hourly and sub-hourly optimization tools like DER-CAM and MATPOWER Optimal Scheduling Tool (MOST) do not consider ELCC either as these tools assume that if the RES contribution is not high enough in any hour of the day, electricity can be purchased from the electricity grid [9] [11] [12]. OSeMOSYS, an open source mixed integer linear programming (MILP) optimization framework, can consider ELCC based on a pre-calculated ELCC curve, which is a function of RES penetration [16]. In the case of this software package, only one RES technology, namely wind, is considered with ELCC. The discussion about ELCC is crucial if an assessment of RES must overcome limited RES and demand data availability or if capacity payments are considered in the energy market of interest.

## III. Methodology

MS-Excel® is used to create a simplified economic merit order dispatch tool in which 50 discretization steps are used instead of an hourly time series for an entire year (8,760 values). This simplification of discretization steps is a common modeling shortcut to speed up calculation while keeping the result close enough (some percent deviation) to results based on 8,760 hours a year [17]. Figure 2 shows the utilized methodology within this paper as a control flow diagram. The start of the considered dynamic programming approach is to calculate the reliability of the existing system without any additions of conventional or renewable generation (step (1)). The system reliability is expressed as LOLE and is based on a time series covering an entire year. The discretization steps are created from a time series of 8,760 hourly values after the load profile has been adjusted for the RES generation within the differently assumed penetrations (net load). Octave (a free and open source MatLab® clone) is called via MS Excel®-Visual Basic (VBA) to calculate the detailed COPT (based on [24]) to improve calculation speed as well as accuracy of the calculations as MS Excel has limitations in terms of large array calculations. In this study, the system reliability metric LOLE is calculated based on the hourly net demand profile of the entire year. After calculating the system reliability, loops around all proposed technology options with capacities between 1 MW and 20 MW (in 1 MW steps) are incorporated (step (2) and (3)). Adding generation capacity to an existing system does improve the system reliability as more generation capacity is available to fulfill the same electricity demand [2]. Adding new generation results in a decreased LOLE. Therefore, step (4) and (5) are required to increase the entire load curve (each hour of the year) to the extent that the system reliability measured as LOLE is the same as it was before adding new generation technologies. The limit for the demand increase is the assessed added name plate capacity. Once the initial LOLE level is reached, the new demand profile is used to be dispatched and assess the economics (step (6)). This procedure is continued as long as any capacity has to be added and technology options have to be assessed.

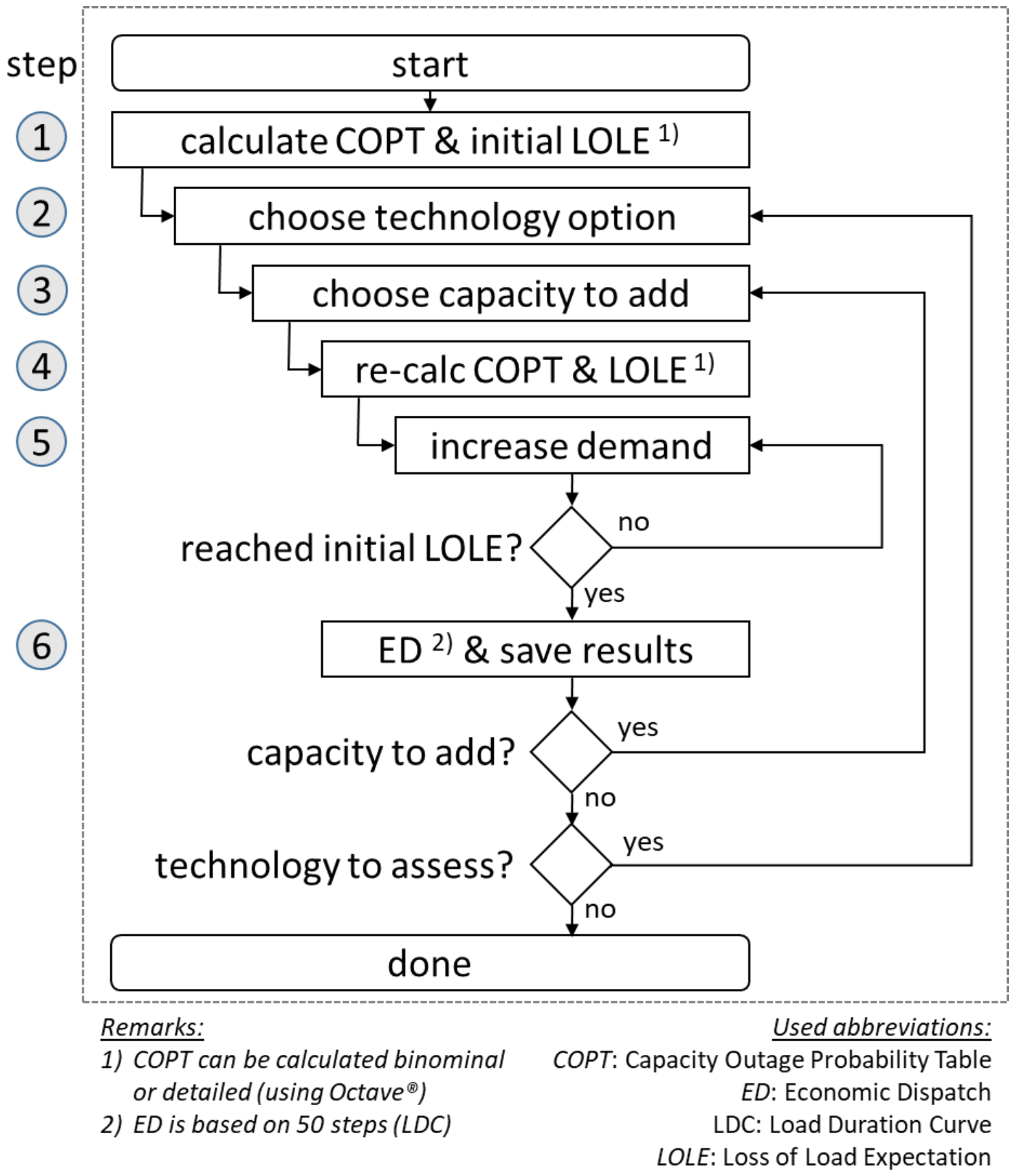

Figure 2: Considered methodology

For dispatch purposes, three different simulation approaches were assessed, basically with different discretization approximations of the load duration curve (LDC). The first approach for the discretization of the LDC starts with the peak value of the expected demand and uses equal size steps to create a LDC step function. The second approach uses each 179$^{th}$ value to create the targeted 50 steps LDC. The third approach uses a k-means cluster analysis to calculate 50 values to minimize the differences between original and calculated values (see equation (1)) [18].

$$argmin = \sum_{i=1}^{k} \sum_{x_j \in S_i} \left\| x_j - \mu_i \right\|^2 \qquad (1)$$

where $k$ is the number of measured data points $x_j$, $j$ is the number of new data points, and $\mu_i$ are the grouped data points.

Option three creates perfect matches within the peak hours and creates average values for the lower demand hours. Figure 3 shows the comparison of the three mentioned approaches to create discretization steps. LDC approaches one and two result in 1% higher energy demand to be fulfilled while the option three comes with the exact energy demand to fulfill. Option two is overestimating the peak hour duration. Option three covers the peak hours perfectly while base load hours are covered on average. Within this work, LDC option three (k-means) is used as it shows the best overall fit.

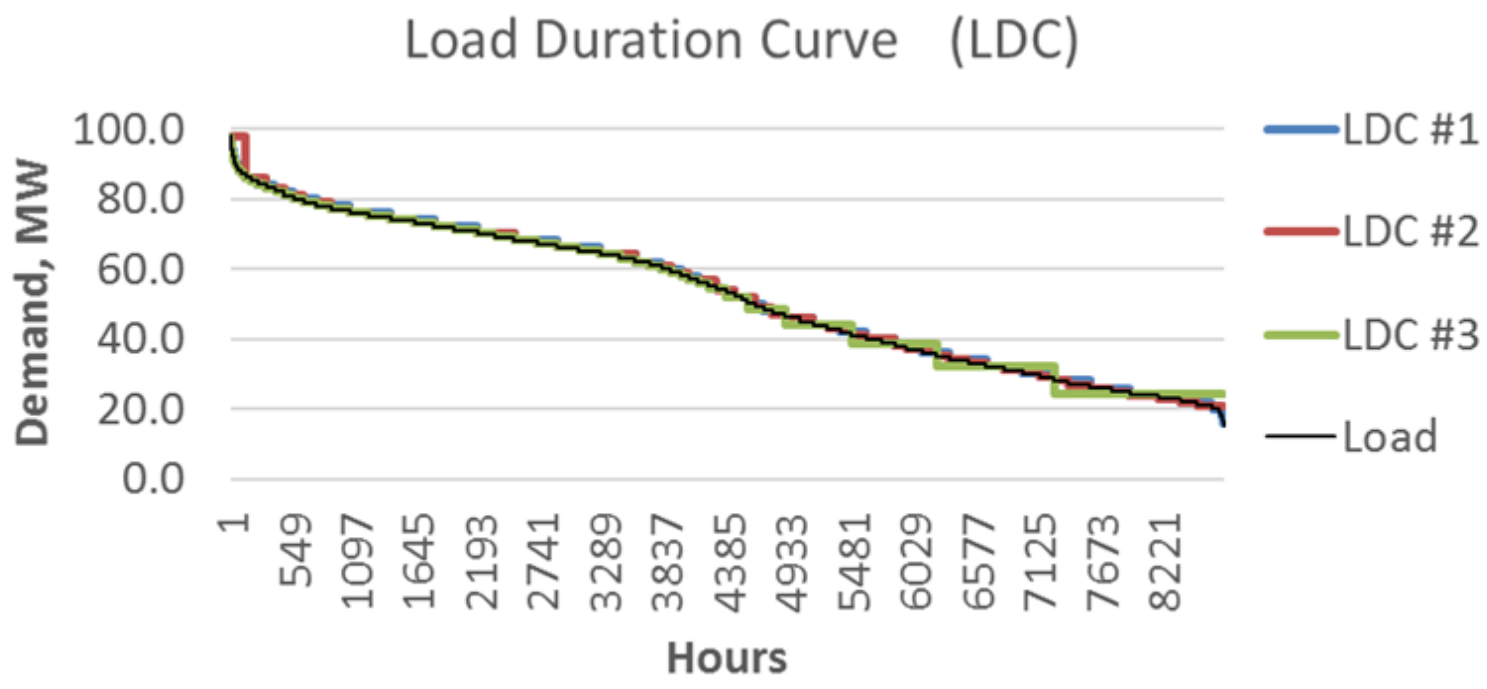

Figure 3: Comparison of different LDC approaches

For each technology option (see step (2) in Figure 2) the economic merit order dispatch is done in which the total annual costs are evaluated based on capital and operational costs considering a total life cycle cost analysis. Capital expenditure annual equivalent values are used to calculate the annualized costs based on the required investment (see equation (2)) [19].

$$A = \left[I - S \cdot \frac{1}{(1+r)^m}\right] \cdot \left[\frac{r \cdot (1+r)^m}{(1+r)^m - 1}\right] \quad (2)$$

where $A$ is the annuity, $I$ is the initial investment cost, $S$ is the salvage value after the operational life time, $r$ is the discount rate, and $m$ represents the operational life time. Within this work, a nominal discount rate of 5%, an inflation of 2%, and zero salvage value has been used. The considered discount rate or weighted average cost of capital (WACC) represents a realistic case where the expected return for the investors' money is 8%, the debt interest rate is 3%, as well as 20% of the investment is from an investor. The WACC does include 1.5% country risk as a premium [20]. The variable operational and maintenance (O&M) costs as well as fuel costs are considered within the simplified dispatch described above where the annuity for new generation (PV, wind, CSP, or open cycle gas turbine (OCGT)) is added to calculate the annual costs.

## IV. Case Study

The system studied is the same system presented in [2], an existing power system isolated from the bulk transmission grid in the southwest area of KSA [21]. Several expansion scenarios are presented targeting the same system reliability (measured in LOLE) as the initial system. These scenarios cover from pure renewable technology expansion scenarios as PV power only at several levels or wind power only at several levels, as well as pure conventional expansion with a OCGT (also known as simple cycle GT, SCGT) or some mix of the mentioned technologies.

### *A. Assumptions*

The current mini-grid consists of a collection of OCGTs: 1x 3 MW unit, 5x 4.2 MW units, 5x 10 MW units, and 3x 17 MW units, resulting in a total installed capacity of 125 MW. Figure 4 shows the shape of LOLP for the mini-grid generation fleet under different FOR assumptions. It is obvious that FOR has a significant impact on the LOLP, and therefore, also on the LOLE. In the present study, a FOR level of 0.05 (or 5%) was chosen for existing as well as added OCGTs as a reasonable assumption, given the real FOR data are not available for the authors. Within the considered hourly VER profiles, FOR is already incorporated accordingly.

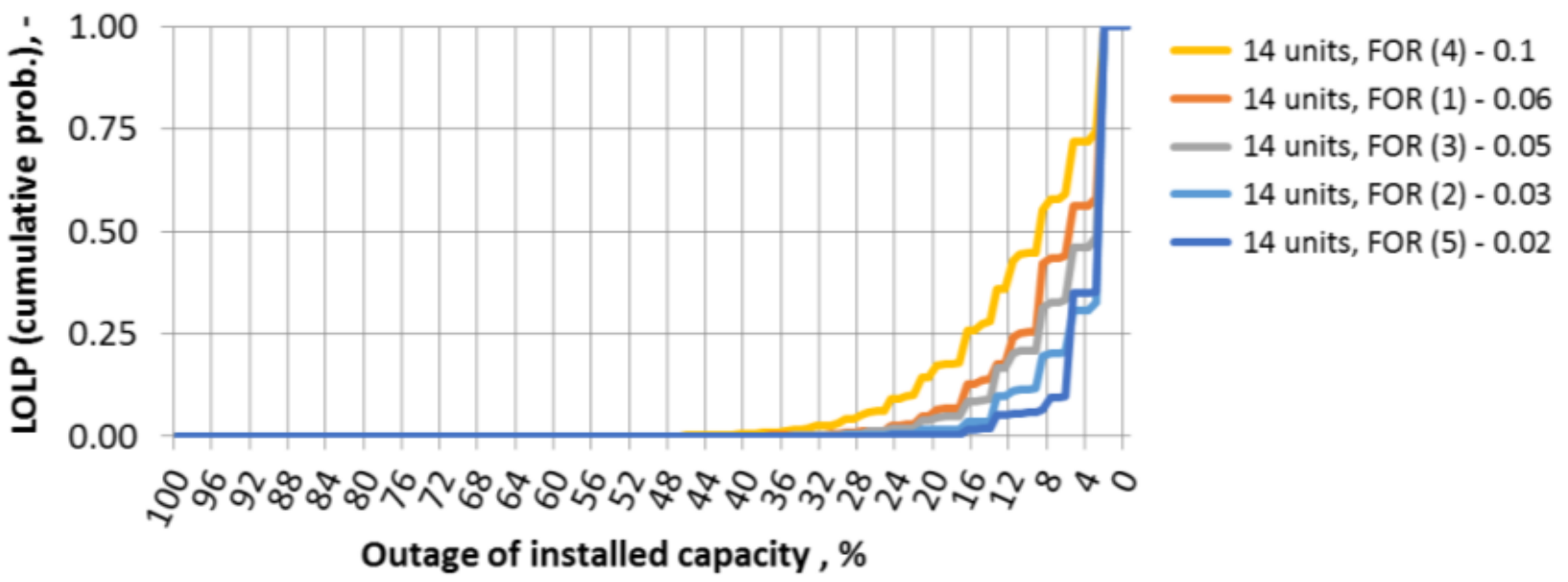

Figure 4: LOLP with 14 units, real sizes and different FOR rates [2]

Considered capital costs (including e. g. land lease, financing, and connection to represent overnight costs) are 1,274 $/kW for ground-mounted PV, 1,632 $/kW for wind, 5,763 $/kW for concentrated solar power (CSP) with 8h storage, and 950 $/kW for OCGT [22]. OCGTs are assumed to operate for 25 years while VER is assumed to run for 20 years only. Fuel costs of $40/barrel crude are assumed where diesel is considered with a 20% premium to cover refining and distribution costs. A sensitivity analysis of ±25% for crude costs, technology life time, and capital costs are considered to assess the importance of the individual parameters. The fact of significant price reductions within renewables, especially PV, in the last years is not required to be considered above the sensitivity analysis as the case study reflects an assessment to install new generation technologies within the next months or so.

LDC option three (k-means) shows the best overall fit to incorporate changes in peak demand as well as not overestimating dispatch costs (see Figure 5). Using the k-mean-based 50-discretization steps results in estimated annual fuel consumption of around 2.3 thousand barrels a day (MBD) while the real fuel demand in 2012 was 2.7 MBD [21]. Considering a constant 10% spinning reserve represents an increase in demand by 10% and results in 2.9 MBD while a constant 10 MW spinning reserve results in 3.0 MBD fuel demand. For this work, the dispatch verification (~±13%) is accurate enough as the real dispatch conditions such as spinning reserve, up/down ramp rates, minimum up/down time, part load efficiency, distribution losses, ambient air conditions [23] are unknown to the authors and are not considered within the selected economic merit order dispatch approach.

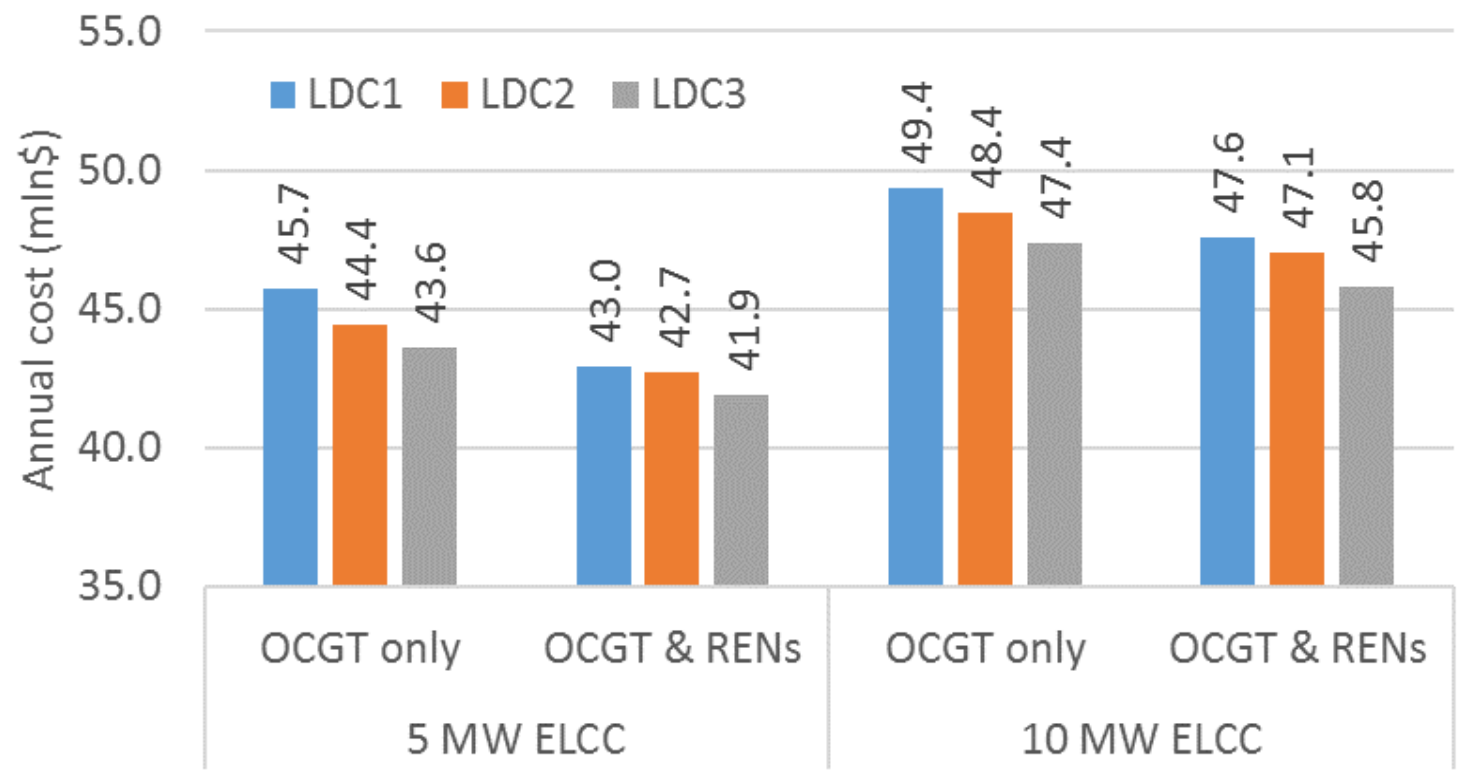

Figure 5: Annual cost comparison for different LDC methods

Figure 6 to Figure 9 show the average day load profile, the average PV output profile (for a fixed tilt system), the average CSP profile, and the average wind profile within the mini-grid location to give an overview of the different output profiles per RES technology [21]. Nevertheless, the detailed calculations like LOLE and ELCC are based on hourly time series and therefore hourly time series were used instead of the shown averaged profiles.

While Figure 6 shows the large difference in summer and winter demand, Figure 7 to Figure 9 do not show significant changes between the seasons. Within this case study, the difference between winter and summer will limit the installation of VER while curtailing renewable energy should be limited for operational and economic reasons.

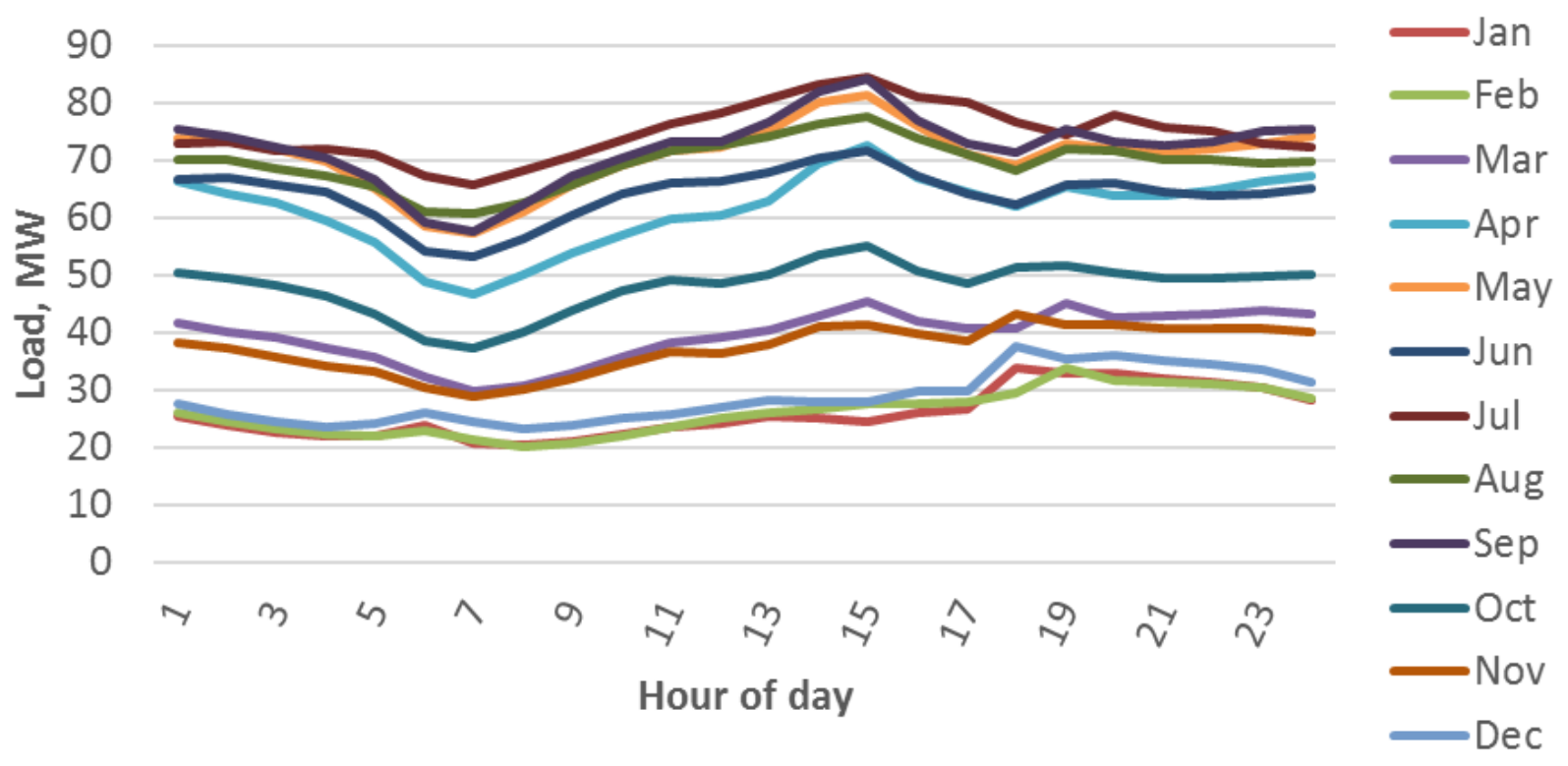

Figure 6: Average day profiles in the mini-grid

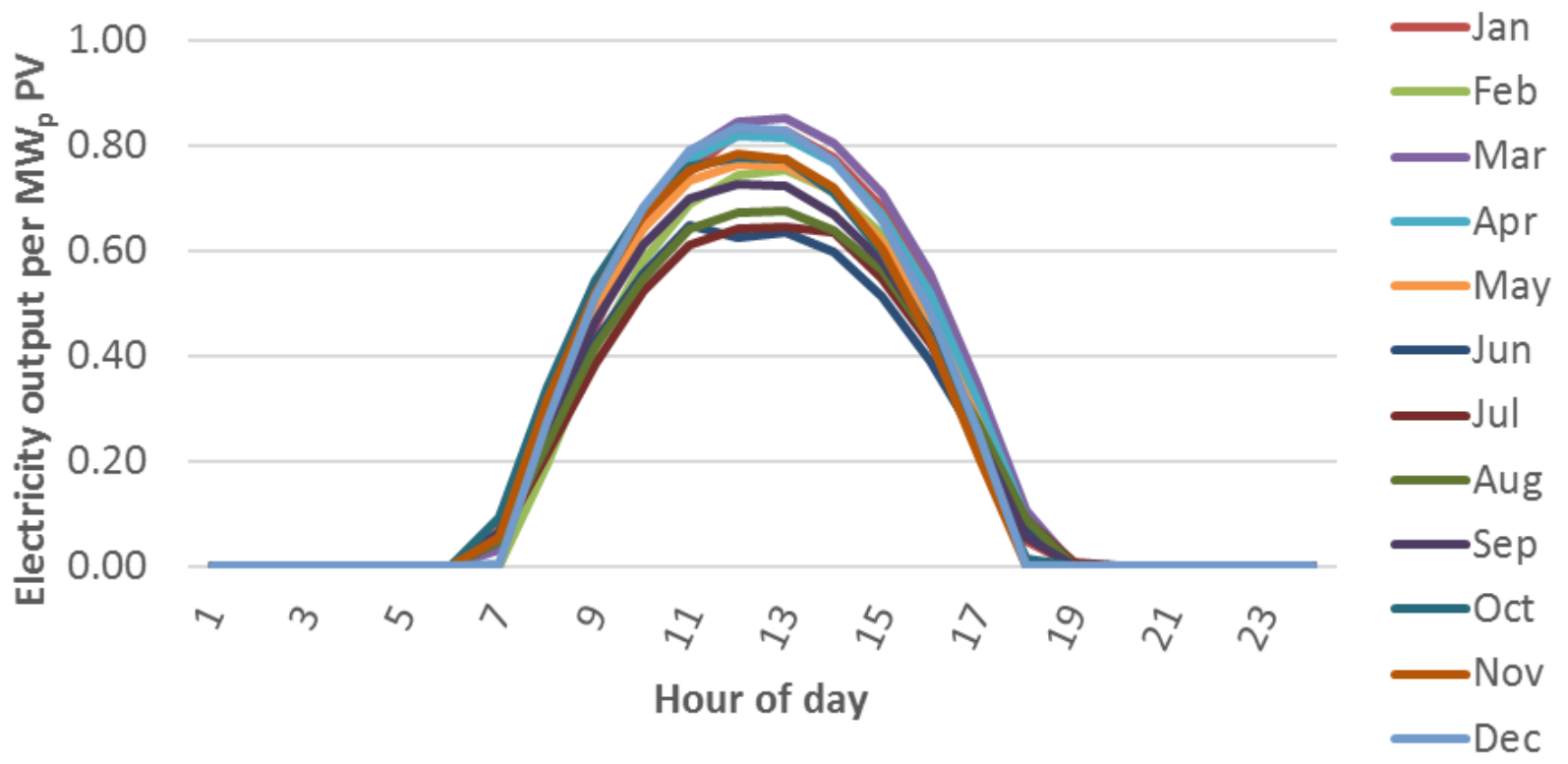

Figure 7: Average PV output profiles in the mini-grid

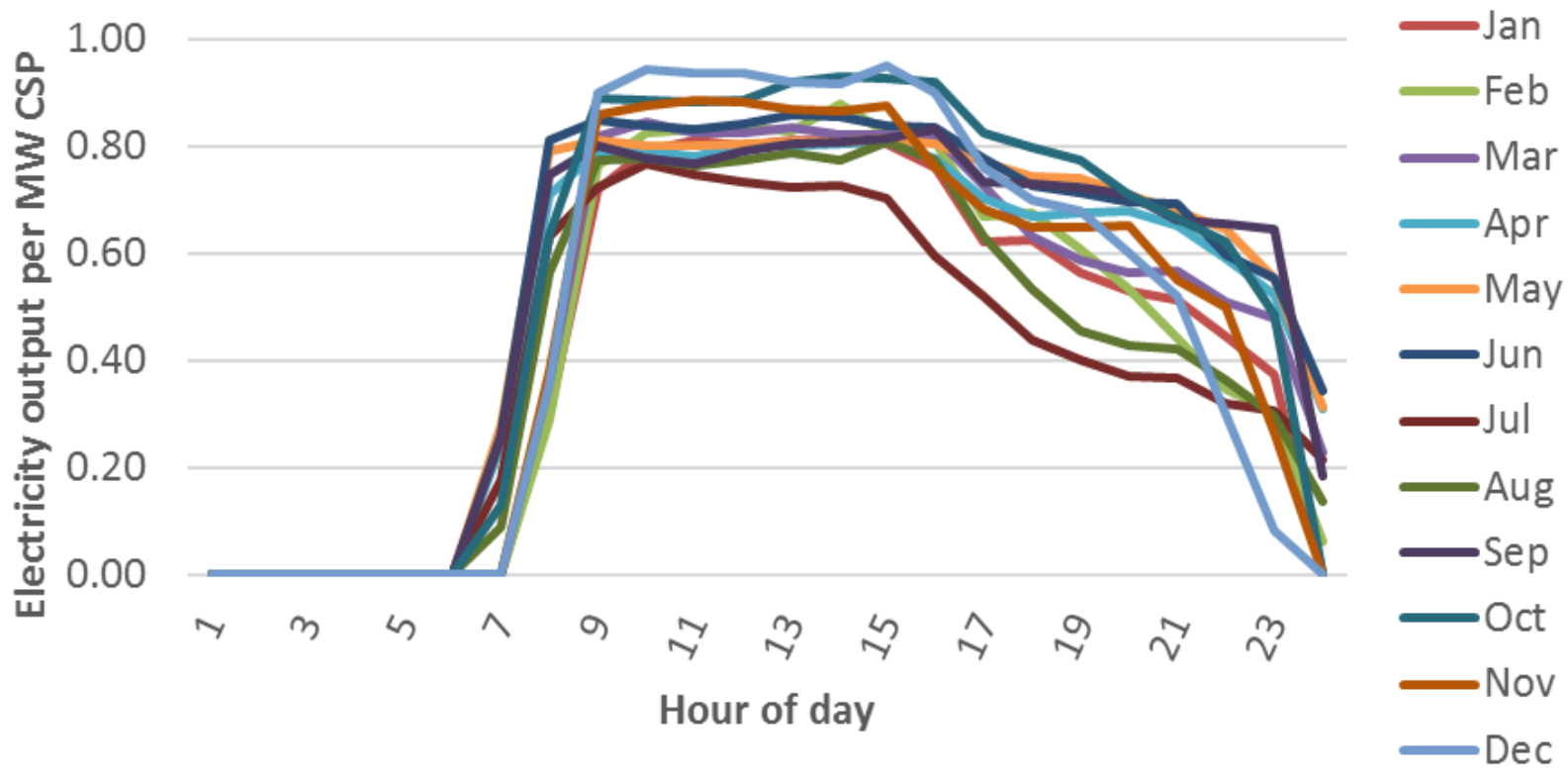

Figure 8: Average CSP output profiles in the mini-grid

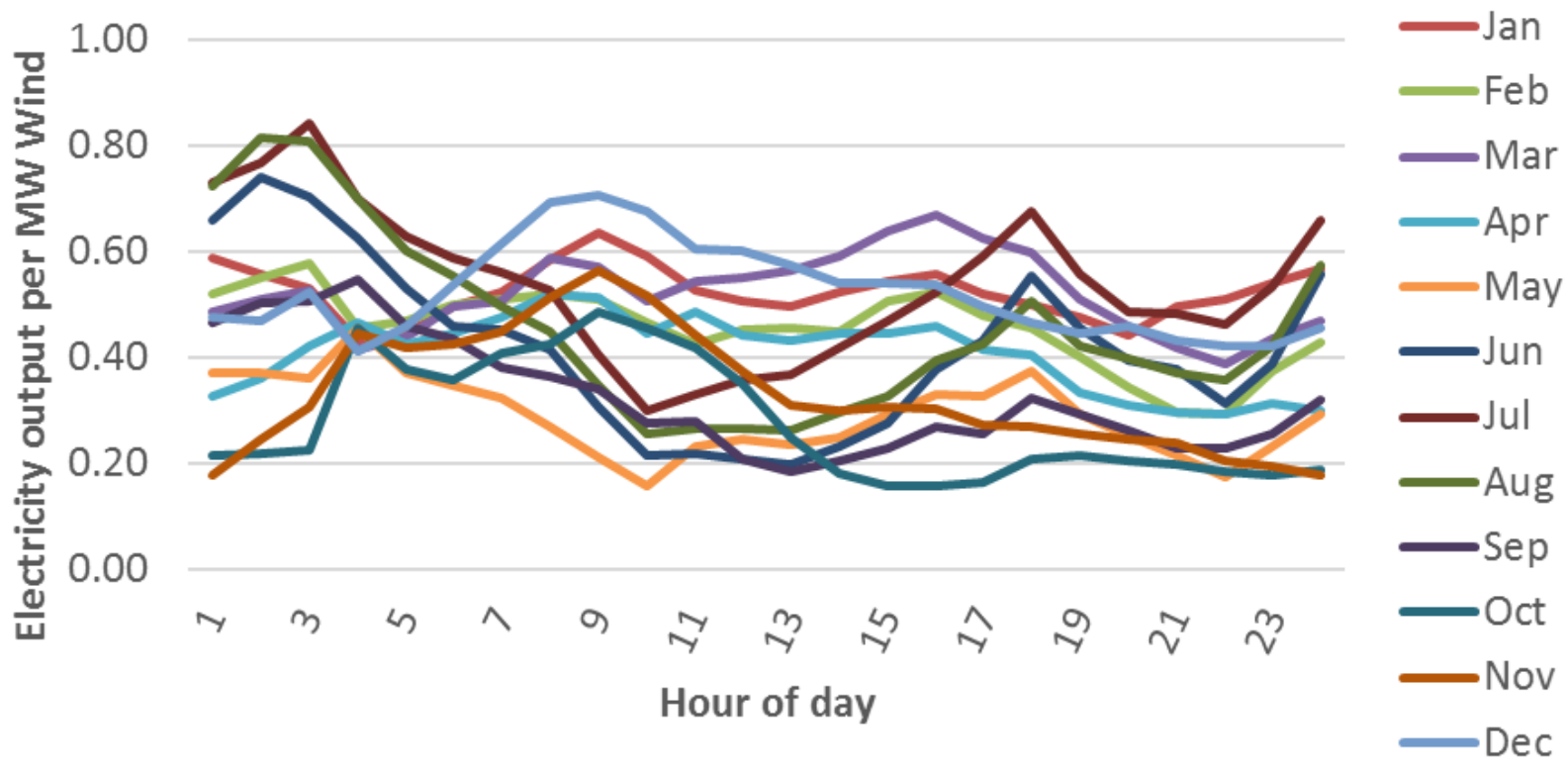

Figure 9: Average wind output profiles in the mini-grid

Table 1 shows the proposed technology options within this work in order to assess the most economic combination of technologies achieving a peak increase of 5% and 10% without decreasing the system reliability (measured in LOLE) of the system.

Table 1: Considered technology options

| technology option | 1 | 2 | 3 | 4 | 5 | 6 | 7 | 8 | 9 | 10 | 11 | 12 | 13 | 14 | 15 | 16 | 17 | 18 | 19 | 20 | 21 | 22 |
|---|---|---|---|---|---|---|---|---|---|---|---|---|---|---|---|---|---|---|---|---|---|---|
| PV share, % | 0 | 100 | 0 | 0 | 75 | 50 | 25 | 60 | 40 | 20 | 20 | 40 | 60 | 15 | 30 | 45 | 10 | 20 | 30 | 5 | 10 | 15 |
| Wind share, % | 0 | 0 | 100 | 0 | 25 | 50 | 75 | 20 | 40 | 60 | 60 | 40 | 20 | 45 | 30 | 15 | 30 | 20 | 10 | 15 | 10 | 5 |
| CSP share, % | 0 | 0 | 0 | 100 | 0 | 0 | 0 | 20 | 20 | 20 | 0 | 0 | 0 | 0 | 0 | 0 | 0 | 0 | 0 | 0 | 0 | 0 |
| OCGT share, % | 100 | 0 | 0 | 0 | 0 | 0 | 0 | 0 | 0 | 0 | 20 | 20 | 20 | 40 | 40 | 40 | 60 | 60 | 60 | 80 | 80 | 80 |
| FOR, % | 5 | 5 | 5 | 5 | 5 | 5 | 5 | 5 | 5 | 5 | 5 | 5 | 5 | 5 | 5 | 5 | 5 | 5 | 5 | 5 | 5 | 5 |

## *B. Results and Discussion*

This chapter describes the results of the case study in which some results are based on reliability measurements only while others are also based on economics.

### *1) Reliability Results*

Figure 10 shows the improvement of the system reliability by adding new generations. For better understanding, LOLE is not shown in logarithmic scale as done usually. It is noticeable that adding capacity, irrespectively of conventional or VER technology, always improves the system's reliability [2]. It is also perceptible that adding more and more capacity for any technology option does not result in a constant improvement. The rate of LOLE enhancement is neither constant nor constantly decreasing as it depends on the status of the system, the already added capacity, and the hourly net demand profile, especially daily demand patterns.

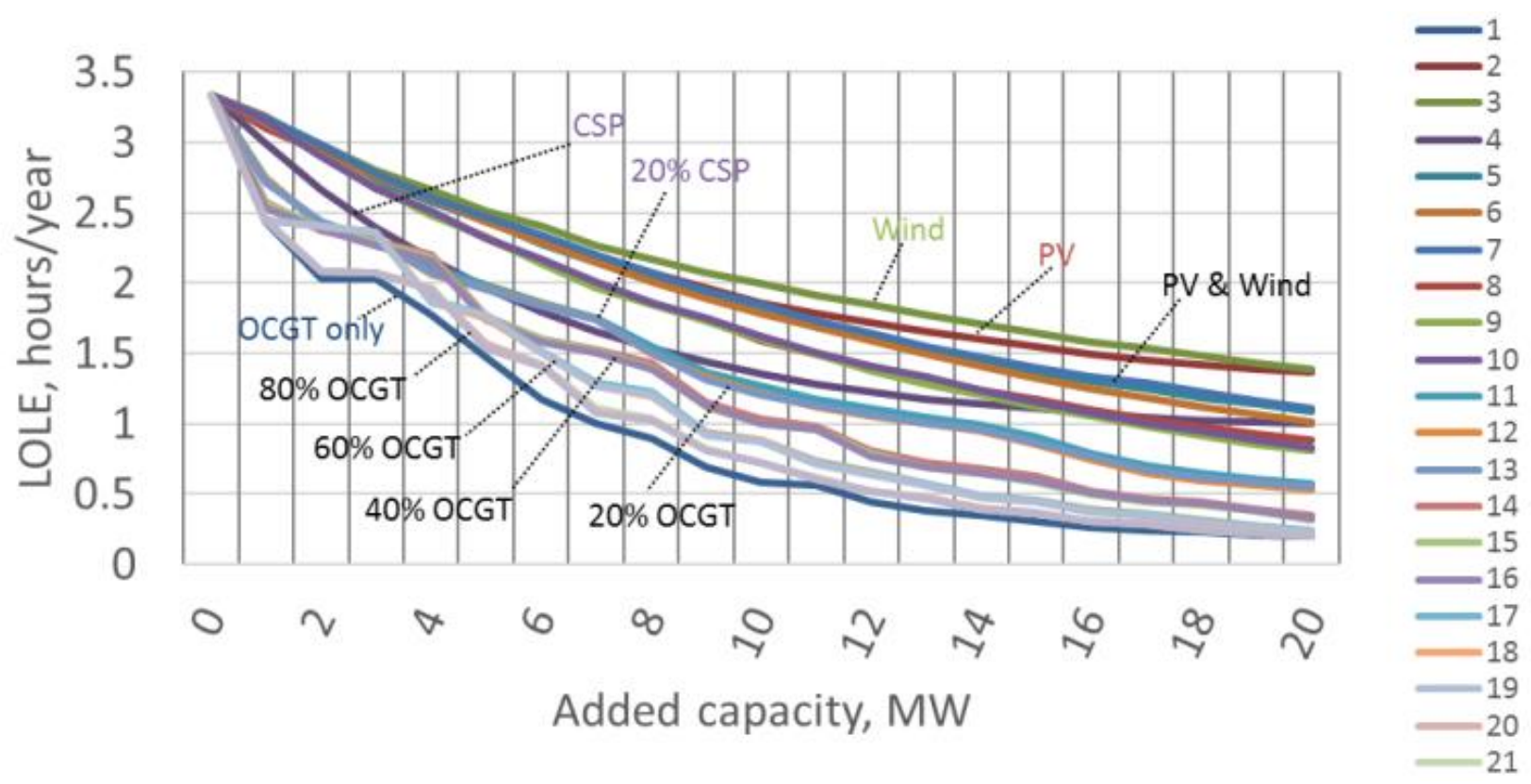

Figure 10: LOLE after capacity additions

Figure 11 shows the ELCC for the considered technology options. The rightmost line represents the OCGT addition showing an addition of 11 MW of capacity results in an ELCC of 10 MW. It is worth mentioning that a purely conventional expansion solution (adding 20 MW of OCGT) results in 14.7 MW ELCC (assuming a FOR of 5% for the new OCGT). Starting with about 7 MW of OCGT addition, the expected ELCC is below the added capacity. Several technology options are not able to reach the required 5 MW or 10 MW ELCC target at all. As an example, the leftmost lines show the addition of PV and wind alone does not result in enough ELCC (around 4 MW of firm capacity) to be a valid result for the requested peak demand increase. The ELCC for an incremental MW (x+1) is always at least the same as for x MW only.

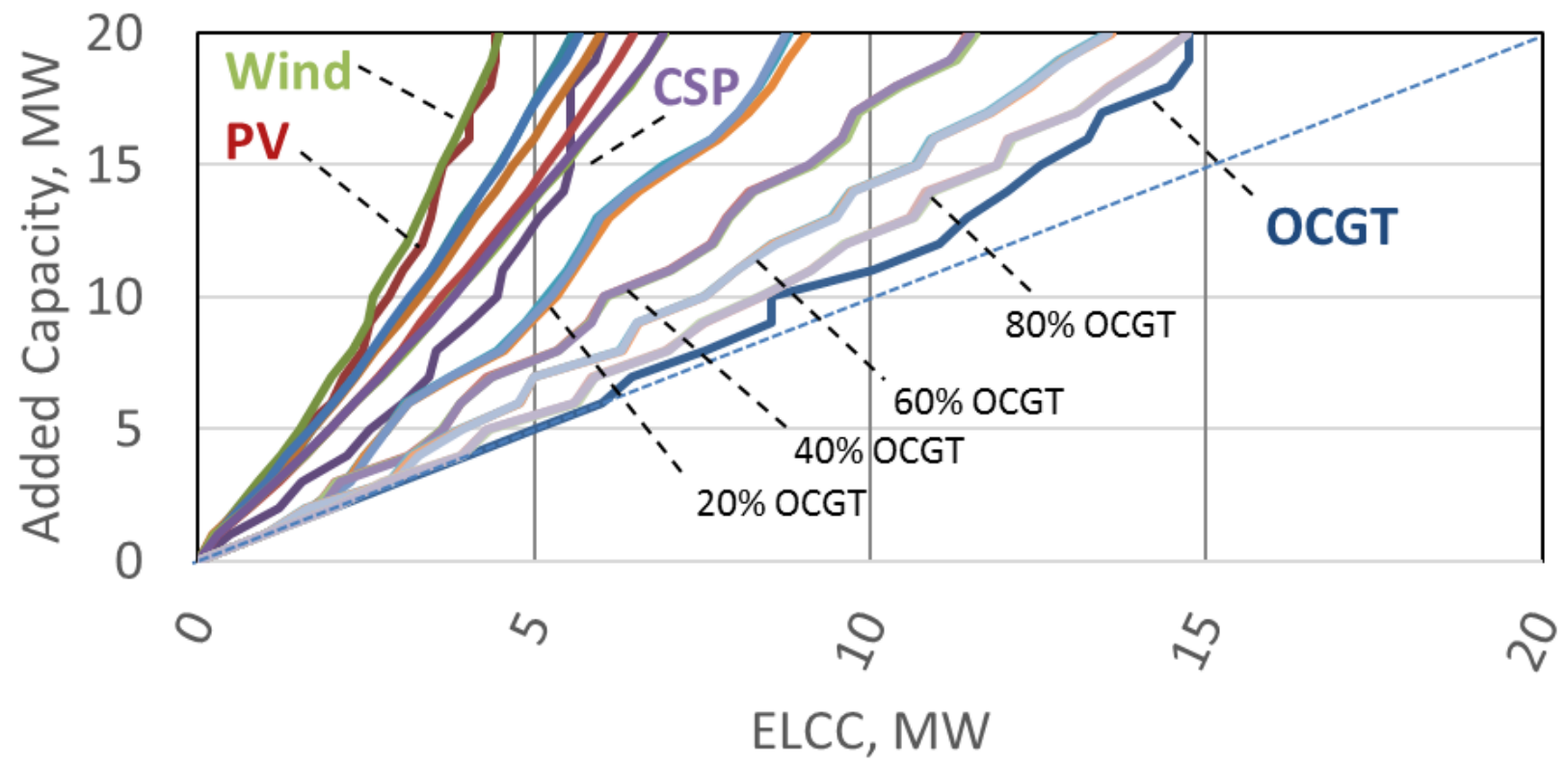

Figure 11: ELCC vs. added capacity

Figure 12 shows the ELCC compared with the expected capacity credit (CC). The CC of any power plant is calculated by dividing the ELCC by the added name plate capacity [2]. Very often ELCC and CC are used interchangeably, which should not be the case. The graph shows that a small addition of new generation has higher capacity value compared with larger additions [2] [4]. The ELCC is calculated after increasing the entire load curve. Therefore, it is important to analyze whether it is a fair approach considering an increase during the day is more likely than during the night. Especially in KSA, an increase during the summer is also more likely than during the winter. The initial increase in CC is a result of the very accurate RES profiles while the demand profile is given in whole MW numbers only as well as the dispatch of discretization steps.

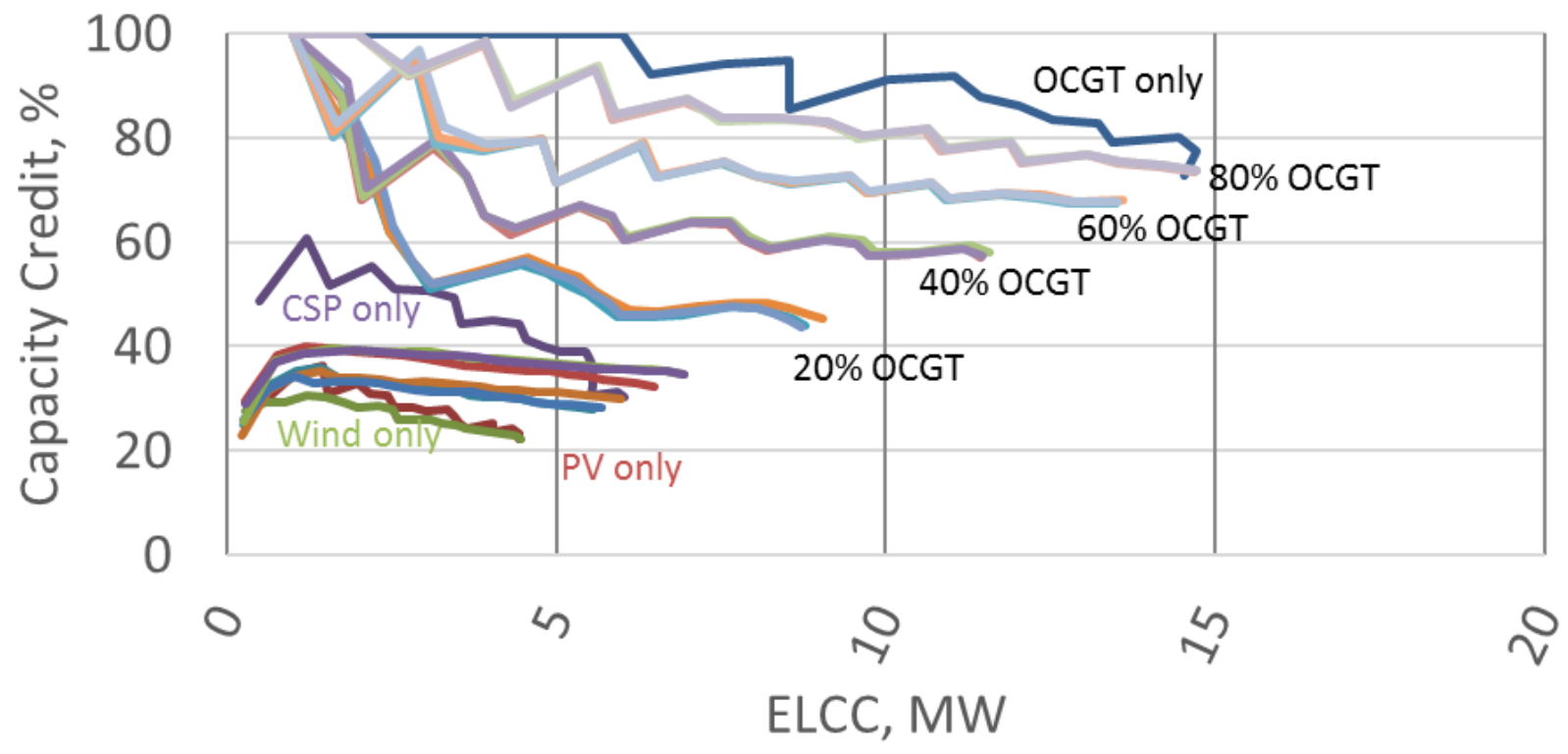

Figure 12: ELCC vs. capacity credit

Figure 13 shows the ELCC results based on daylight hours and the entire day. The expected ELCC results for daylight are not the same as the ELCC for the entire day. While wind has almost the same CC, the results are quite different for PV and CSP. This is because wind is quite volatile and shows low correlation with the demand profile while PV and CSP have a very high correlation with the daily demand pattern (Figure 6 to Figure 9). Another reason is that PV is considered with no storage and CSP is considered with 8h storage only. Consequently, PV is not able to move enough energy into the night hours (between 7 and 9 p.m.) in which the second peak of the day occurs. These two shown approaches display significant differences, especially for sun-driven technologies as PV and CSP. Therefore, it is important to define, from an electricity market perspective, how RES should be treated to ensure security of the electricity supply. Different approaches are shown in [25] (e.g. peak time only, summer month's only). One approach could be that the daylight LOLE approach is used as long as the net demand shows a peak demand during the day and once the peak demand is after sunset the daylong LOLE approach could be considered. It is important to remember that the ELCC results from other countries have to be taken with care as this metric is very sensitive to demand and available VER resources.

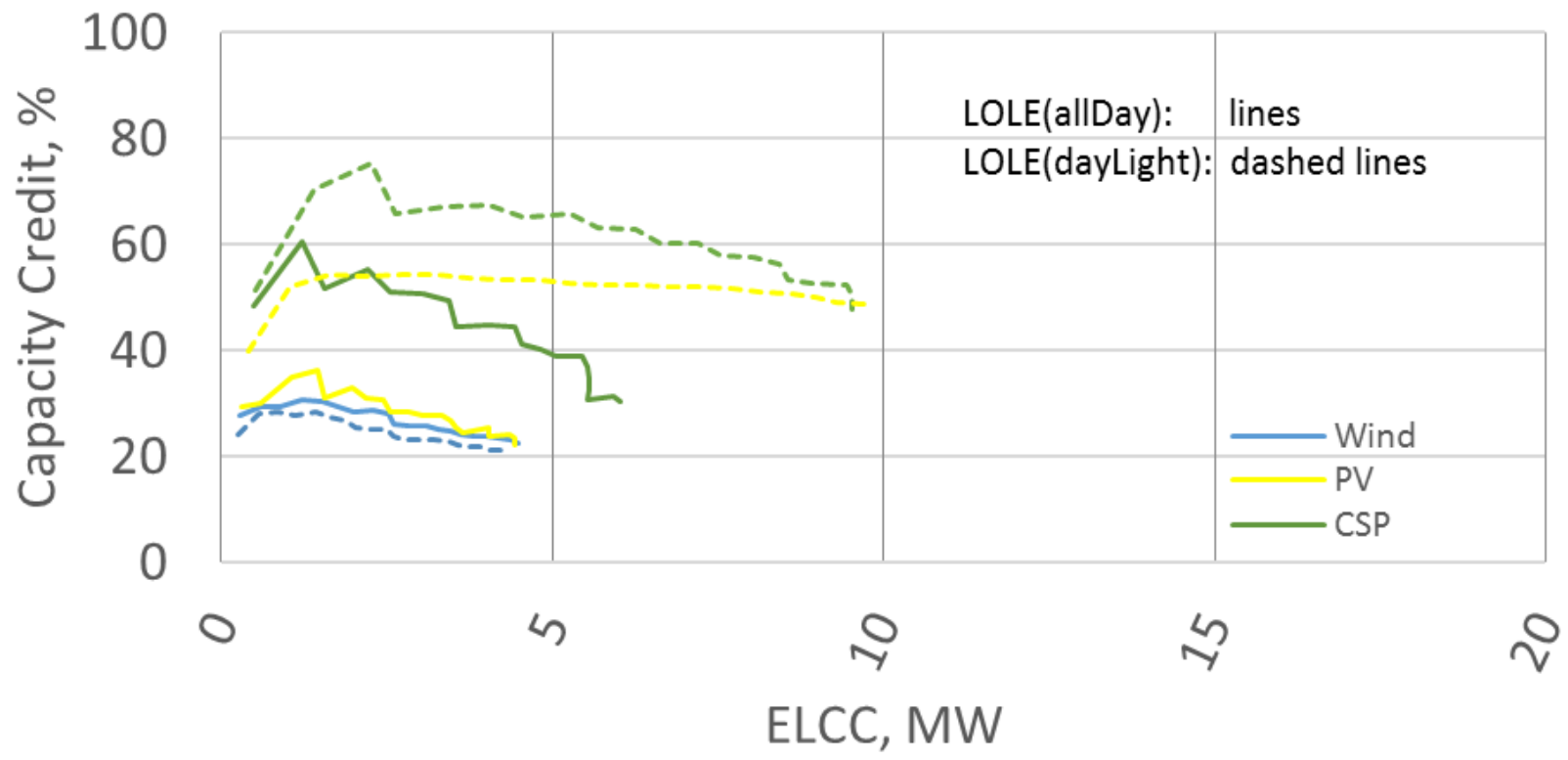

Figure 13: ELCC vs. capacity credit for pure RES options

Figure 14 shows the ELCC for pure RES options as well as RES combinations with different FOR assumptions. Mixing different technologies can have a positive impact on ELCC as well as CC. For example, the ELCC for PV and wind combined is bigger than for wind alone. The combined RES technology lines also show flatter CC curves. As a result, for wind it is valuable to add PV as this significantly increases the ELCC as well as the expected CC.

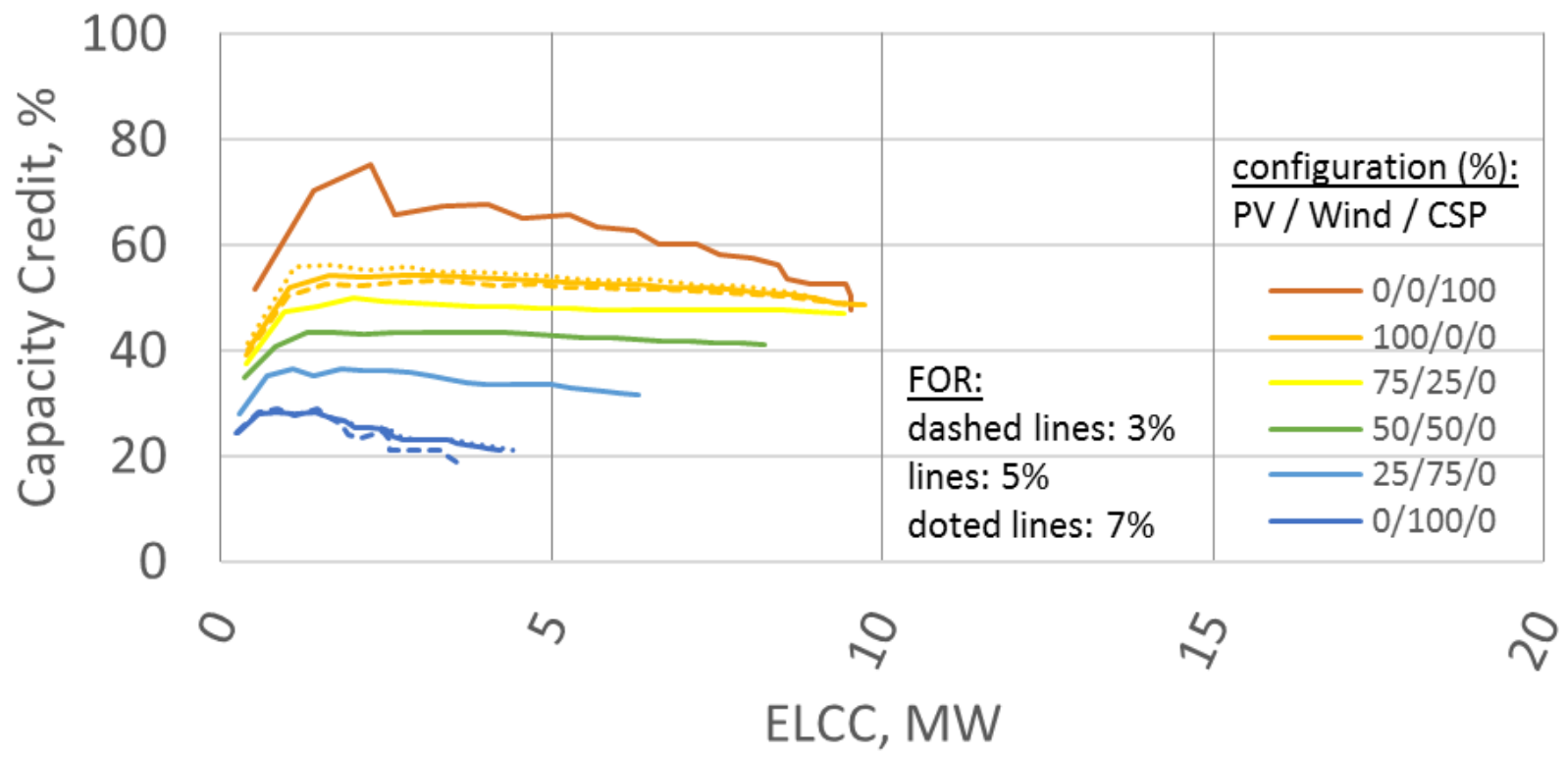

Figure 14: ELCC vs. capacity credit for RES combinations (FOR sensitivity)

Figure 15 shows the renewable shares of electricity generation from all the options of RES and OCGT for each ELCC value required (the dark blue line from OCGT shows, of course, a zero-renewable energy component). A solution scenario adding 100% wind shows the highest energy contribution of VER with 76 GWh (18% of the annual power demand within the given time series), but fails to achieve the required ELCC of 5 MW or 10 MW being able to serve a maximum ELCC, or peak load, of 4.5 MW. Additional wind has quite low ELCC but contributes with low-cost energy, and therefore, could and should be considered besides the ELCC calculations. Also, PV can significantly contribute to fuel savings even if ELCC is limited. In other words, even the ELCC-based decision would favor OCGT solutions additions of VER do result in significant fuel savings. A finding worth mentioning is that within the 5 MW ELCC solution the combination of 9.6 MW wind, 3.2 MW PV, and 3.2 MW CSP is just 2% (or $0.64 million) more expensive than the least cost combination while it is 3% (or $1.23 million) cheaper than the OCGT only solution. With higher shares of CSP (above the 20% share considered in this work) the CSP options might be included in the 10 MW ELCC solutions.

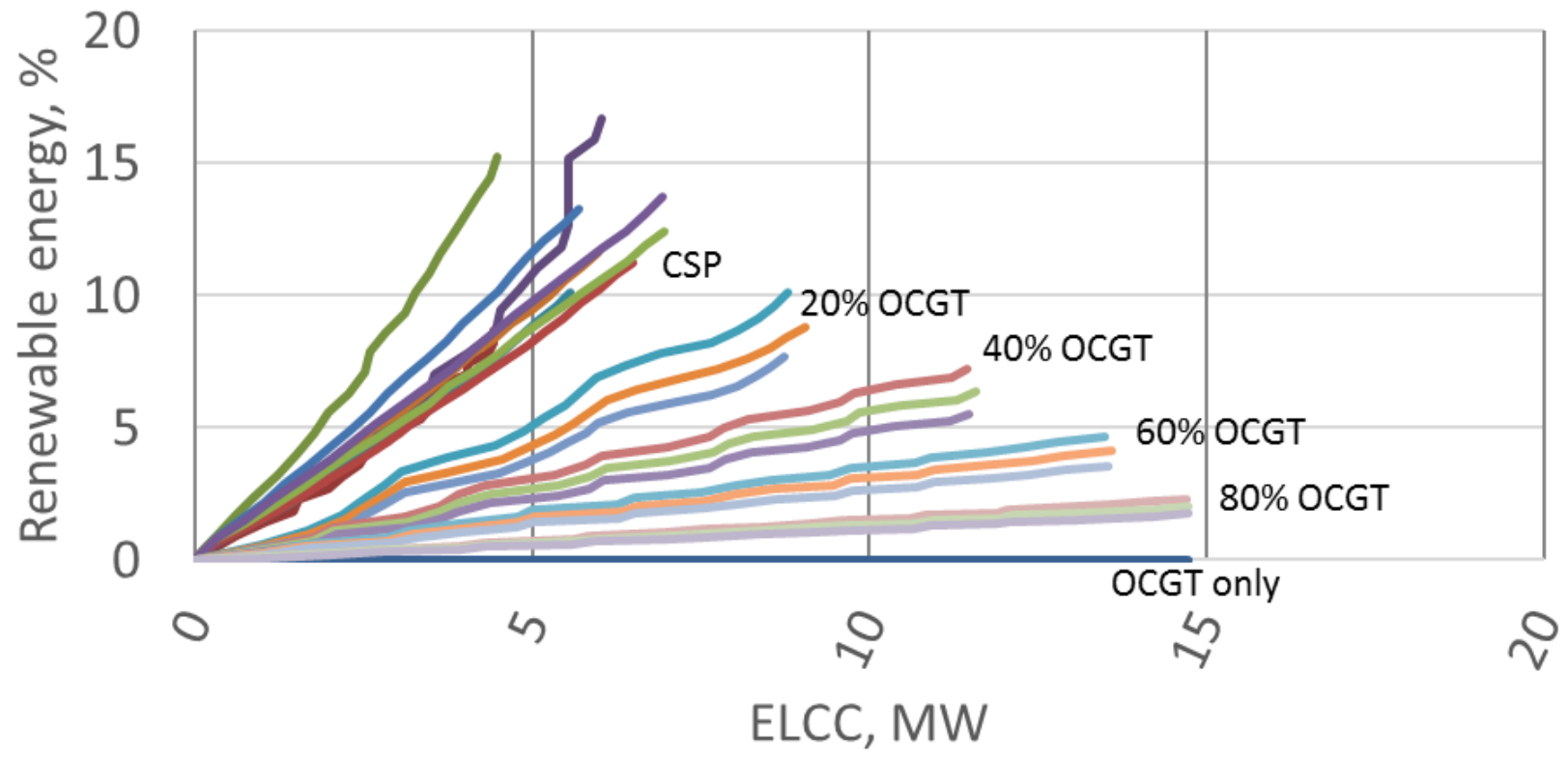

Figure 15: ELCC vs. renewable energy produced

Until now, all shown results are independent of assumed fuel prices as they are based on system reliability.

### 2) *Economic Results*

Table 2 and Table 3 show all feasible combinations able to reach required peak demand increase (or ELCC) of 5 MW (or 5%) and 10 MW (or 10%), respectively, with crude prices of $30, $40, and $50/barrel. An increase in fuel prices results in more viable options to be competitive within the 10 MW ELCC cases in terms of economics as fuel spending can be further reduced. Within the 5 MW ELCC cases (see Table 2), the $30 and $40/barrel scenarios favor 25% PV and 75% wind, the $50/barrel scenario favors 60% PV, 20% wind, and 20% CSP. The shift to CSP is the result of the thermal energy storage (TES) assumed within the CSP system and as fuel is expensive enough, CSP would be an economic choice to replace some conventional generators after sunset. Both, CSP and wind, are a form of base load technology and hence in direct competition. Once CSP is found more economical than wind, it would reduce added wind capacity and shift the wind capacity towards PV to reduce peak demand. The 5 MW ELCC case does not need any conventional generation. This is because the electricity provided from RES technologies is more expensive than the annualized payment to install the RES technology and also able to allow a 5 MW peak increase. Within the 10 MW ELCC cases (see Table 3), all fuel price scenarios result in the most economic technology combination of 15% PV, 45% wind, and 40% OCGT. While relatively small increments of ELCC (up to 5 MW or 5%) can be realized with RES only, higher ELCC increments require OCGT to be included, based on the considered reliability constraints. Higher fuel costs allow more VER options to be economically viable. Most valid options, including the RES only options, result in lower annual costs than the OCGT only solution. This is because a fuel price of $30/barrel or more is assumed to be constant for the entire operational life of all generation technologies and is high enough to pay back (through fuel savings) the additional RES capital requirements.

Table 2: Total annual cost for valid 5% peak increase

| **technology option** | **1** | **4** | **5** | **6** | **7** | **8** | **9** | **10** | **11** | **12** | **13** | **14** | **19** | **20** |
|---|---|---|---|---|---|---|---|---|---|---|---|---|---|---|
| PV share, % | 0 | 0 | **75** | **50** | **25** | **60** | **40** | **20** | **20** | **40** | **60** | **15** | **30** | **5** |
| Wind share, % | 0 | 0 | **25** | **50** | **75** | **20** | **40** | **60** | **60** | **40** | **20** | **45** | **10** | **15** |
| CSP share, % | 0 | **100** | 0 | 0 | 0 | **20** | **20** | **20** | 0 | 0 | 0 | 0 | 0 | 0 |
| OCGT share, % | **100** | 0 | 0 | 0 | 0 | 0 | 0 | 0 | **20** | **20** | **20** | **40** | **60** | **80** |
| added cap. (MW) | 5 | 19 | 18 | 20 | 19 | 17 | 14 | 16 | 13 | 11 | 13 | 8 | 7 | 7 |
| | **30$/barrel** | | | | | | | | | | | | | |
| 5 MW ELCC | 33.4 | | 32.4 | 32.4 | **31.2** | 31.4 | 33.4 | 33.1 | 33.1 | | | 33.4 | 33.4 | |
| ( = option - OCGT only ) | | | 2.9% | 3.1% | 6.9% | 6.3% | 0.0% | 0.9% | 0.9% | | | -0.1% | 0.1% | |
| | **40$/barrel** | | | | | | | | | | | | | |
| 5 MW ELCC | 43.6 | | 41.9 | 41.7 | **40.1** | **40.1** | 42.9 | 42.3 | 42.5 | 42.8 | 43.4 | 43.5 | 43.5 | 43.5 |
| ( = option - OCGT only ) | | | 4.1% | 4.5% | 8.6% | 8.6% | 1.6% | 2.9% | 2.6% | 1.8% | 0.5% | 0.2% | 0.3% | 0.1% |
| | **50$/barrel** | | | | | | | | | | | | | |
| 5 MW ELCC | 53.8 | 53.4 | 51.3 | 51.0 | 49.0 | **48.8** | 52.4 | 51.6 | 52.1 | 52.6 | 53.3 | 53.5 | 53.6 | 53.7 |
| ( = option - OCGT only ) | | 0.8% | 4.9% | 5.5% | 9.7% | 10.1% | 2.6% | 4.2% | 3.1% | 2.2% | 0.9% | 0.4% | 0.4% | 0.2% |
| | | | | | | | | | | | | | | |
| **std.dev** | 10.2 | | 9.4 | 9.3 | 8.9 | **8.7** | 9.5 | 9.3 | 9.5 | 6.9 | 7.0 | 10.1 | 10.1 | 7.2 |

Table 3: Total annual cost for valid 10% peak increase

| **technology option** | **1** | **14** | **15** | **16** | **17** | **18** | **19** | **20** | **21** |
|---|---|---|---|---|---|---|---|---|---|
| PV share, % | 0 | **15** | **30** | **45** | **10** | **20** | **30** | **5** | **10** |
| Wind share, % | 0 | **45** | **30** | **15** | **30** | **20** | **10** | **15** | **10** |
| CSP share, % | 0 | 0 | 0 | 0 | 0 | 0 | 0 | 0 | 0 |
| OCGT share, % | **100** | **40** | **40** | **40** | **60** | **60** | **60** | **80** | **80** |
| added cap. (MW) | 11 | 18 | 18 | 18 | 15 | 15 | 15 | 13 | 14 |
| | **30$/barrel** | | | | | | | | |
| 10 MW ELCC | 36.4 | **35.4** | 35.8 | 36.3 | | | 36.4 | 36.1 | 36.2 |
| ( = option - OCGT only ) | | 2.7% | 1.7% | 0.3% | | | 0.0% | 0.8% | 0.6% |
| | **40$/barrel** | | | | | | | | |
| 10 MW ELCC | 47.4 | **45.8** | 46.3 | 47.0 | | 47.4 | 47.3 | 46.9 | 47.0 |
| ( = option - OCGT only ) | | 3.5% | 2.4% | 0.9% | | 0.1% | 0.3% | 1.1% | 0.8% |
| | **50$/barrel** | | | | | | | | |
| 10 MW ELCC | 58.4 | **56.2** | 56.8 | 57.7 | 58.5 | 58.2 | 58.1 | 57.7 | 57.8 |
| ( = option - OCGT only ) | | 4.0% | 2.8% | 1.3% | -0.1% | 0.3% | 0.5% | 1.2% | 1.0% |
| | | | | | | | | | |
| **std.dev** | 11.0 | 10.4 | 10.5 | 10.7 | | **7.7** | 10.8 | 10.8 | 10.8 |

Figure 16 shows an option analysis for the least cost option within the 10 MW ELCC case. This analysis considers adjusted parameters with ±25%. Oil price is by far the most sensitive factor for the annualized total costs. An oil price change by ±25% shows a change of annual costs of ±22% while all other parameters are showing changes lower than ±1%. For the 5 MW ELCC OCGT only solution, the standard deviation of annual costs is $10.2 million while for the RES only solution it is $8.6 million on average (TABLE 2 and 3). For the 10 MW ELCC OCGT only solution, the standard deviation is $11.0 million while for solutions including RES it is $10.2 million. Therefore, RES can be used to hedge against fuel price increases and minimize financial burdens associated with that. It is also visible that technology life time and capital expenditure for new generation technologies do not have a significant impact on the result as the annuities of each of them is quite low compared with the fuel cost within the studied mini-grid.

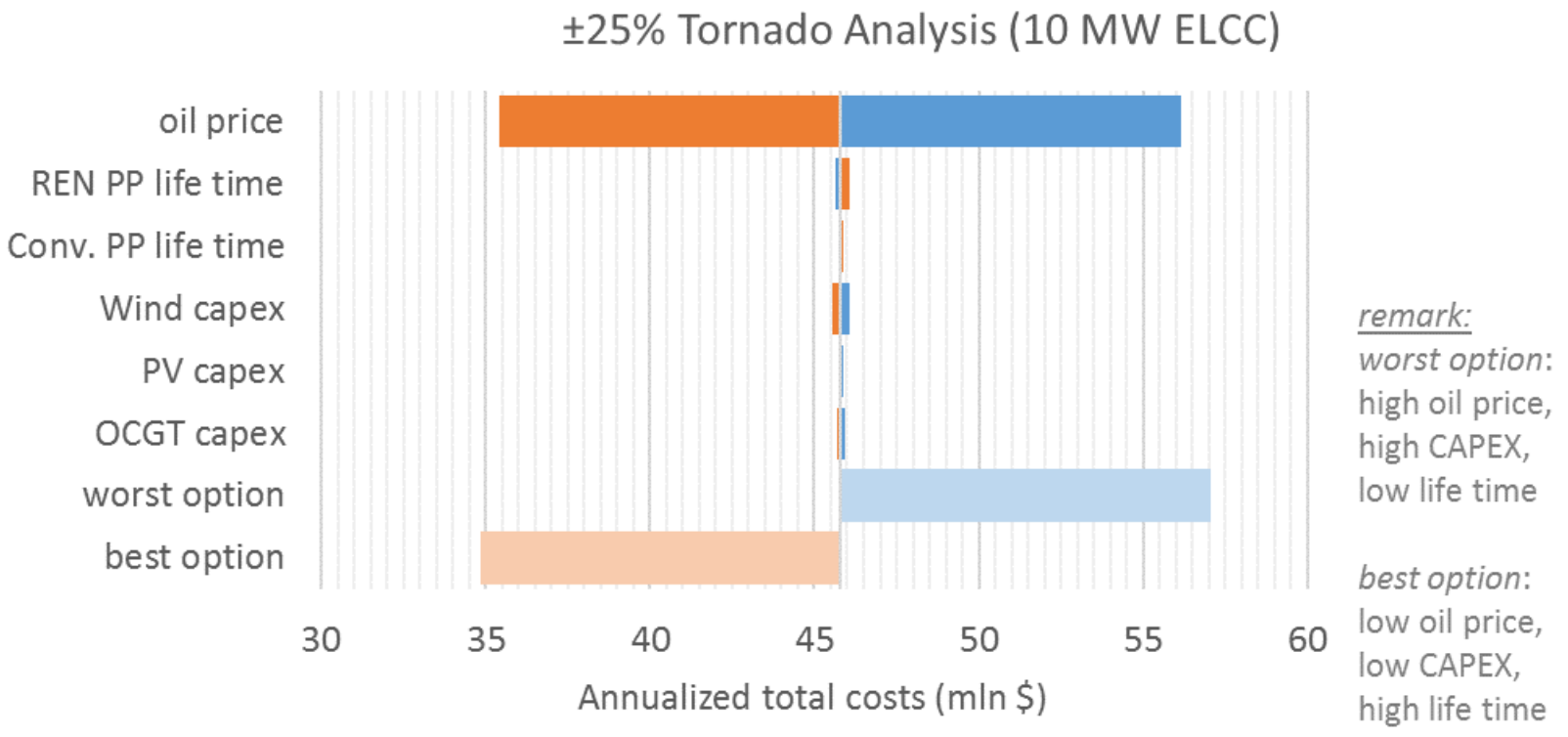

Figure 16: Tornado analysis for key input assumptions (10 MW ELCC)

Figure 17 shows that considering the RES only solutions has higher potentials to decrease the financial burden compared with the OCGT only solution. Within the considered and predefined technology combinations, the potentially lowest cost solution would be the 25%/75% mix of PV and wind, respectively.

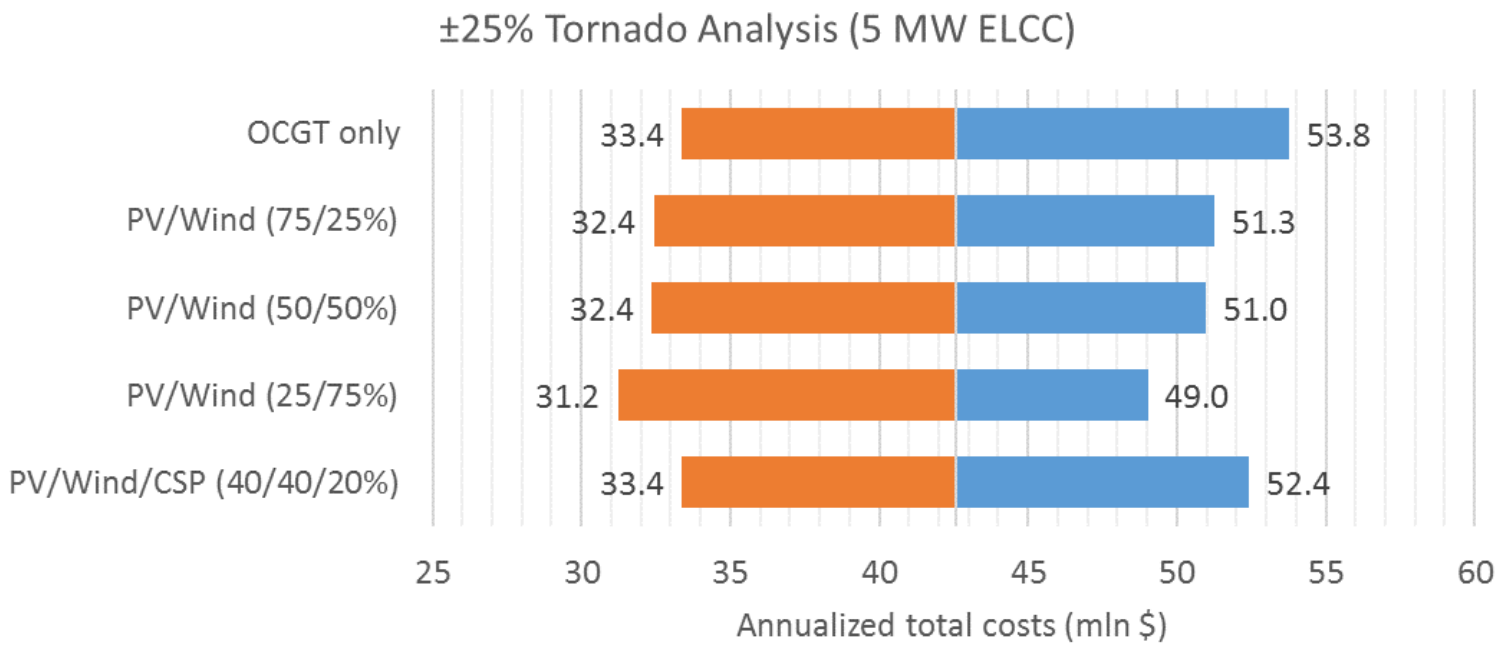

Figure 17: Fuel price sensitivity (5 MW ELCC)

A limitation of this case study is that only one year of data is available and therefore the results have to be considered indicative only. Multiple years of data should be used to get a more accurate understanding of the RES profile volatility. A disadvantage of the ELCC procedure is that the entire demand curve is increased while in reality there might be more increase during the day than during the night hours. To overcome this, an hourly bottom-up demand forecast or using the multi-seasonal exponential smoothing method could be used to create more realistic results [26].

## V. Conclusions

With the introduction of VER into GEP, more probabilistic approaches are necessary to ensure proper planning and a reliable power system. Therefore, concepts considering COPT, LOLP, and ELCC are used to plan for reliable electricity systems. If a marginal growth of up to 5 MW (or 5%) is expected within a short period of time, it is more economical to install further RES capacity as the available capacity is enough to cover critical hours in which RES is not available. On the other hand, if a more significant growth (in our case 10%) is expected within a 'short' time of period, it is more favorable to add conventional generation as well as RES allowing peak demand to grow and to limit or reduce fuel demand growth. The results show significant shares of conventional generation technologies. Nevertheless, from an economic perspective, additional RES generation should be added for all considered fuel price scenarios (> $30/barrel), because of fuel-saving reason alone. Figure 11 shows that most VER options have lower annual costs than OCGT only solutions. Based on the considered fuel prices and especially in remote areas, where the mini-grid within this case study is located, the use of CSP is a viable and economical option for continuous operation ("base load") as well as controllable (dispatch-able) generation capacity. Both, PV and wind provide very economical contributions for fuel displacement and programs.

## VI. Future Work

Further research may incorporate PV tracker systems into the assessment and study their impact on the mini-grid economics. Additionally, using an explicit socioeconomic cost for interrupted power may also give a different perspective on the technology options for the expanded system. Another interesting topic could be to enhance the proposed methodology by considering power flow details (with an open source tool as MATPOWER or PSAT [27]) and other dispatch constraints (e.g. up/down ramp rates, and minimum up/down time [23]) to optimize the hourly or sub-hourly dispatch problem. Also, the use of different sizes of storage as well as curtailment of RES provided electricity could be assessed within the given system reliability and economic constraints, within the next work.

## Acknowledgments

The authors gratefully acknowledge valuable comments and support from colleagues and management within Saudi Aramco (especially from Power Systems Renewables Department and Power Systems Planning Department) as well as from anonymous reviewers.

## Biographies

**Markus Groissböck** is a doctoral student of Energy Systems Engineering at the Institute for Construction and Materials Science, Unit of Energy Efficient Buildings, University of Innsbruck. In 2009, he earned a Bachelor degree in Energy Economics from the University of Applied Sciences Kufstein, Austria, as well as a Master degree from the University of Applied Sciences Burgenland, Austria, and a Master degree from the University of Liverpool in Sustainable Energy Systems and Operations and Supply Chain Management (Oil and Gas), UK, in 2011 and 2016, respectively. He has been a Power Generation Expansion Planner for about 10 years. His research interests lie in medium- and long-term investment planning within the power and energy sector considering distributed and volatile energy resources on district and regional level.

**Alexandre Gusmao** has been a Renewable Energy Business Developer for more than 15 years, covering a portfolio of projects from wind, PV, hydro and biomass. His interests are in planning for renewable integration in power grid and mini-grids, forecasting natural energy resources, forecasting assets prices.